\documentclass[12pt,a4paper]{article}

\usepackage[T1]{fontenc}
\usepackage{lmodern}
\usepackage{amsmath,amssymb,amsthm,mathtools}
\usepackage{geometry}
\usepackage{url}
\usepackage{microtype}
\usepackage{graphicx}
\usepackage{booktabs}
\usepackage[hidelinks,bookmarks=false]{hyperref}

\numberwithin{equation}{section}

\newcommand{\email}[1]{\texttt{#1}}
\newcommand{\address}[1]{\date{#1}}
\newcommand{\keywords}[1]{\par\medskip\noindent\textbf{Keywords: }#1\par}
\newcommand{\refcite}[1]{\cite{#1}}

\newcommand{\supp}{\operatorname{supp}}
\newcommand{\dist}{\operatorname{dist}}
\newcommand{\Tr}{\operatorname{Tr}}

\newcommand{\E}{\mathbb E}
\newcommand{\R}{\mathbb R}
\newcommand{\C}{\mathbb C}
\newcommand{\Z}{\mathbb Z}

\newcommand{\cD}{\mathcal D}
\newcommand{\cG}{\mathcal G}
\newcommand{\cA}{\mathcal A}
\newcommand{\Om}{\Omega}
\newcommand{\ellength}{\operatorname{length}}

\renewcommand{\Re}{\operatorname{Re}}
\renewcommand{\Im}{\operatorname{Im}}

\newtheorem{thm}{Theorem}
\newtheorem{lem}[thm]{Lemma}

\newtheorem{prop}[thm]{Proposition}
\newtheorem{cor}[thm]{Corollary}

\newtheorem{ass}[thm]{Assumption}
\newtheorem{rem}[thm]{Remark}

\begin{document}

\markboth{M. Kaminaga}{Atomic limit and analytic DOS in a random Kronig-Penney model}
\title{Controlled atomic limit and analytic density of states in a strongly attractive random Kronig-Penney model}
\author{Masahiro Kaminaga}
\address{Department of Information Technology, Tohoku Gakuin University,\\
3-1 Shimizu-koji, Wakabayashi-ku, Sendai-shi, Miyagi 984-8588, Japan\\
\email{kaminaga@g.tohoku-gakuin.ac.jp}}
\maketitle

\begin{abstract}
We study the negative-energy density of states of a one-dimensional random
Kronig-Penney model with independent and identically distributed attractive
delta interactions.  Each isolated interaction of strength $q<0$ has a bound
state of energy $-q^2/4$.  Using the inverse coupling $V=-1/q$, the Krein
resolvent formula reduces the continuum problem to a lattice operator with
random diagonal entries and energy-dependent hopping.  Under a local
analyticity assumption on the coupling density, we prove a controlled atomic
limit.  After energy is divided by the square of the attraction strength,
the scaled density of states approaches the density of isolated binding
energies with an exponentially small error on compact interior intervals.
An explicit strong-attraction condition also gives real-analyticity of the
density of states and the integrated density of states on the corresponding
negative-energy interval.  A support condition ensures that this interval
belongs to the almost-sure spectrum.  For a uniform coupling law, we obtain a
concrete sufficient threshold and illustrate the spectral profile with
finite-volume calculations based on the exact point-interaction recurrence.
\end{abstract}


\section{Introduction}\label{sec:intro}

The density of states (DOS) is one of the basic spectral quantities used to
characterize disordered quantum systems.  In one dimension, randomness in the
scattering strengths redistributes spectral weight and can substantially
modify its energy dependence.  The random Kronig-Penney model provides a
simple continuum setting in which this effect can be studied without
replacing the scatterers by regular potentials.  In the strongly attractive
regime, each scatterer supports a deep local bound state, while propagation
between distinct centers gives a weak hybridization of these local states.
The model therefore describes the effects of random local binding and
propagation between centers in a disordered continuum system.

We consider the one-dimensional random point-interaction Hamiltonian
\begin{equation}
H_\omega=-\frac{d^2}{d x^2}
 +\sum_{n\in\Z}q_\omega(n)\,\delta(x-n)
\quad\hbox{in }L^2(\R),
\label{formal-H}
\end{equation}
Here the lattice spacing is the unit of length and $\hbar^2/(2m)=1$.
The random variables $q_\omega(n)$ are independent and identically
distributed.  We assume throughout the paper that the interactions are
attractive and uniformly separated from zero, so that their common
distribution is supported in a compact subset of $(-\infty,0)$.  For a single
attractive delta interaction of strength $q<0$, the bound-state energy is
\begin{equation}
E_{\rm b}(q)=-\frac{q^2}{4}.
\label{single-bound-state}
\end{equation}
Thus, under the scaling $q\mapsto\lambda q$, the natural negative-energy scale
is $E=O(\lambda^2)$.  The energy interval considered below follows this same
single-scatterer binding-energy scale.

Basic results on one-dimensional point interactions can be found in
Refs.~\refcite{AlbeverioEtAl1984,Albeverio2005}.  Spectral questions for random
point defects were studied in Ref.~\refcite{AlbeverioHKM1982}.  For the
one-dimensional random Kronig-Penney model, Drabkin, Kirsch, and
Schulz-Baldes~\cite{DrabkinKirschSchulzBaldes2012} analyzed transport and DOS
singularities near positive critical energies.  Kirsch and
Nitzschner~\cite{KirschNitzschner1990} studied IDS asymptotics at band edges,
while Hislop, Kirsch, and Krishna~\cite{HislopKirschKrishna2020} proved
Poisson local eigenvalue statistics in a negative-energy localization regime.
The relation between random spectra and periodic comparison operators is
also a general theme for random potentials; see
Ref.~\refcite{KirschMartinelli1982}.

The ergodic trace formulation of the IDS has its background in
Refs.~\refcite{Pastur1980,Shubin1979}; the latter concerns almost-periodic
elliptic operators.  For stochastic Jacobi matrices, Craig and
Simon~\cite{CraigSimon1983} proved log-H\"older continuity of the IDS.
Campanino and Klein~\cite{CampaninoKlein1986} obtained higher differentiability
for the one-dimensional Anderson model under regularity conditions on the
single-site distribution.  Analyticity at strong disorder was proved by
Bellissard and Hislop~\cite{BellissardHislop2007} for densities analytic in a
strip, and local analyticity was treated by Kaminaga, Krishna, and
Nakamura~\cite{KaminagaKrishnaNakamura2012}.  The negative-energy interior
window considered here is distinct from the positive critical energies and
band-edge regimes mentioned above.

The physical question addressed here is how the negative-energy spectral
profile is organized when local binding dominates propagation, and how stable
this profile is when the weak coupling between different centers is restored.
Our main results give a controlled atomic limit and, on the same energy
window, a real-analytic disorder-averaged DOS.  More precisely, under a
quantitative strong-attraction condition and a local analyticity assumption on
the coupling distribution, the density-of-states measure has a real-analytic
density on a finite interval of negative energy, and the IDS is real-analytic
there.  We further prove, under a natural support condition, that this
interval is contained in the almost-sure spectrum.  Thus the result applies to an interval inside the random spectrum.

The continuum problem is not a direct instance of the discrete Anderson model.
To see the difference, introduce the inverse coupling
\begin{equation}
V_\omega(n):=-\frac{1}{q_\omega(n)}>0.
\label{intro-inverse-coupling}
\end{equation}
At energy $z=-\kappa^2$, the Krein resolvent formula reduces the continuum
problem to an operator on $\ell^2(\Z)$ whose diagonal entries contain
$V_\omega(n)-(2\kappa)^{-1}$.  The effective spectral parameter is therefore
\begin{equation}
\zeta=\frac{1}{2\kappa}.
\label{intro-zeta}
\end{equation}
At the same time, the off-diagonal matrix elements depend on the energy and
are proportional to
$\zeta\exp(-|n-m|/(2\zeta))$.  Thus the spectral parameter enters both the
random diagonal part and the hopping term.  Moreover, the natural random
variable in the reduced problem is not the original coupling $q_\omega(n)$
but its inverse $V_\omega(n)$.  These features distinguish the present
continuum problem from the usual discrete Anderson Hamiltonian.

In the strong-attraction regime, $\zeta=O(\lambda^{-1})$, and propagation
between distinct scatterers is exponentially suppressed.  The Krein
reduction therefore produces an Anderson-type lattice problem with random
diagonal terms and exponentially weak, energy-dependent long-range hopping.
This observation also explains why the relevant negative-energy scale is
$O(\lambda^2)$.  In addition to proving regularity of the DOS, we identify an
interval on this physical binding-energy scale and show that it lies inside
the almost-sure spectrum.  More quantitatively, we prove that the rescaled DOS
converges exponentially to the probability density obtained by mapping the
single-site law through $q\mapsto-q^2/4$.  This gives a controlled description
of the negative-energy spectral weight in terms of random local binding
energies, with an error estimate for propagation between centers.

The proof uses the random-walk and complex-deformation argument of
Ref.~\refcite{KaminagaKrishnaNakamura2012} for the discrete Anderson model, but
several additional steps are required in the present continuum setting.  The
Krein formula must first be used to obtain the effective lattice operator, the
inverse-coupling variable has to be introduced, and the energy-dependent
long-range hopping must be controlled uniformly in a complex neighborhood.
After these reductions, disorder averaging factorizes the contributions
associated with repeated visits of a random walk to the same site.

An important point is that the analytic continuation is performed only after
the disorder average is taken.  On a real energy interval inside the random
spectrum, the factors $V_\omega(n)-\zeta$ cannot be assumed to remain
uniformly separated from zero.  Such a pointwise condition would remove the
relevant values of the random variable and could place the corresponding
energy in a spectral gap.  Instead, we begin with $\Im\zeta>0$, where a
Neumann expansion is valid, average each path over the disorder, and then
analytically continue the resulting single-site integrals across the real
$\zeta$ axis.  This procedure yields a real-analytic disorder-averaged DOS
without imposing a pointwise resolvent bound on the real axis.

The strong-attraction condition is also quantitative.  For the scaled
couplings $q_\omega^{(\lambda)}(n)=\lambda\widetilde q_\omega(n)$, the
off-diagonal propagation is exponentially small in $\lambda$, while the
single-site bounds grow only algebraically.  Consequently, the convergence
condition is automatically satisfied for sufficiently large $\lambda$.  For
the uniform distribution considered in the numerical example, we obtain an
explicit sufficient threshold and compare the corresponding analytic interval
with a finite-volume DOS calculated directly from the exact point-interaction
recurrence.  The numerical calculation is not used to establish analyticity;
its purpose is to show where the rigorous interval is located in the spectral
profile.

The result concerns the regularity of the disorder-averaged DOS.  It should
not be interpreted as a statement about extended states or quantum transport.
In particular, analyticity of the DOS does not by itself determine the
localization length or the conductivity.  The result instead identifies a
strong-binding regime in which a singular one-dimensional continuum system
has a highly regular disorder-averaged spectral profile, even on an interval
belonging to the random spectrum.

The paper is organized as follows.  Section~\ref{sec:model} defines the model,
the IDS, and the inverse-coupling distribution.  Section~\ref{sec:krein}
derives the effective lattice description in the strong-binding regime.
Section~\ref{sec:single-site} introduces the local disorder averages needed in
the binding-energy window.  Section~\ref{sec:rwe} relates weak propagation
between centers to analyticity of the DOS.  Section~\ref{sec:spectrum} locates the
analytic interval in the almost-sure spectrum and proves the controlled
atomic-limit approximation.  Section~\ref{sec:numerics} gives a finite-volume
numerical illustration, and Section~\ref{sec:discussion} summarizes the
physical interpretation and limitations.  Technical continuation estimates are collected in Appendix~\ref{app:continuation}.

\section{The random point-interaction model}\label{sec:model}
%%%%%%%%%%%%%%%%%%%%%%%%%%%%%%%%%%%%%%%%%%%%%%%%%%%%%%%%%%%%

Let $\{q_\omega(n)\}_{n\in\Z}$ be i.i.d. real random variables.  We assume
that there are constants
$$
-\infty<q_-<q_+<0
$$
such that
$$
\supp\mu_q\subset[q_-,q_+],
$$
where $\mu_q$ is their common law and $\supp$ denotes support.

The Hamiltonian $H_\omega$ is defined by
the closed lower-bounded quadratic form
\begin{equation}
h_\omega[u]
 =\int_\R |u'(x)|^2\,d x
  +\sum_{n\in\Z}q_\omega(n)|u(n)|^2,
\qquad
\cD(h_\omega)=H^1(\R).
\label{form}
\end{equation}
Here $\cD(h_\omega)$ denotes the form domain of $h_\omega$.  We always use
the continuous representative of a function in $H^1(\R)$, so that the point
values $u(n)$ in \eqref{form} are well defined.  The standard sampling estimate
for $H^1(\R)$ implies that the second term is form bounded with arbitrarily
small relative bound.  Hence \eqref{form} defines a self-adjoint lower-bounded
operator.  More precisely, its operator domain consists of the functions
$u\in H^1(\R)$ whose restrictions to $(n,n+1)$ belong to $H^2(n,n+1)$, whose
second derivatives form an $L^2(\R)$ function away from $\Z$, which are
continuous across every interaction point,
$u(n-0)=u(n+0)=u(n)$ for $n\in\Z$, and which satisfy the standard
point-interaction jump condition
\begin{equation}
u'(n+0)-u'(n-0)=q_\omega(n)u(n).
\label{jump}
\end{equation}
This quadratic-form construction and the corresponding domain
characterization are standard for one-dimensional point interactions;
see Refs.~\refcite{AlbeverioEtAl1984,Albeverio2005}.
The family is ergodic under integer translations.  For $L>0$, let
$H_{\omega,L}$ denote the Dirichlet restriction to $(-L,L)$.  The integrated
density of states (IDS) is defined by
\begin{equation}
N(E)=\lim_{L\to\infty}\frac{1}{2L}
\#\{\lambda\le E:\lambda\hbox{ is an eigenvalue of }H_{\omega,L}\},
\label{IDSdef}
\end{equation}
Here $\#$ denotes cardinality.  The limit exists at continuity points almost surely and is
nonrandom.  We take $N$ to be right-continuous.  Equivalently, with
$C=[0,1)$, let $\chi_C$ denote multiplication by the characteristic function
of $C$, and let $\chi_B(H_\omega)$ denote the spectral projection of
$H_\omega$ onto a Borel set $B$.  We write $\E$ for expectation over the
disorder and $\Tr$ for the trace.  The ergodic trace formula (usually called
the Pastur-Shubin formula; see Refs.~\refcite{Pastur1980,Shubin1979} for its background)
then gives
\begin{equation}
N(E)=\E\Bigl[\Tr\bigl(\chi_C\chi_{(-\infty,E]}(H_\omega)\chi_C\bigr)\Bigr].
\label{PS}
\end{equation}
Let $n$ be the density-of-states measure determined by
$N(E)=n((-\infty,E])$.  When this measure is absolutely continuous on an
interval, we use the same symbol $n(E)$ for its density there.  For
$z\in\C_+:=\{z:\Im z>0\}$ put
\begin{equation}
m_+(z)=\E\Bigl[\Tr\bigl(\chi_C(H_\omega-z)^{-1}\chi_C\bigr)\Bigr].
\label{mplus}
\end{equation}
In one dimension the localized resolvent is trace class, with trace norm
locally bounded in $z\in\C_+$ uniformly in $\omega$.  Indeed, the uniform form
bound implies $H_\omega+c_0\ge c_1(-d^2/dx^2+1)$ for suitable $c_0,c_1>0$.
The localized inverse of the operator on the right has finite trace.
Factoring the resolvent through $(H_\omega+c_0)^{-1/2}$ then gives the stated
trace-norm bound.  The ergodic trace formula and the spectral theorem give
\begin{equation}
m_+(z)=\int_\R\frac{d n(\lambda)}{\lambda-z}.
\label{Stieltjes}
\end{equation}
Consequently, on an interval where the DOS measure is absolutely continuous,
\begin{equation}
n(E)=\frac{1}{\pi}\lim_{\epsilon\downarrow0}
\Im m_+(E+i\epsilon)
\label{DOSformula}
\end{equation}
holds for Lebesgue almost every $E$ in that interval.  Under the hypotheses
of Theorem~\ref{thm:main}, the density has a real-analytic representative
and the limit in \eqref{DOSformula} equals this representative at every point
of the interval.

We now introduce the inverse coupling variable
\begin{equation}
V_\omega(n)=-\frac{1}{q_\omega(n)}.
\label{inverse-coupling}
\end{equation}
Its common law is denoted by $\nu$.  Since $q_\omega(n)<0$, the support of
$\nu$ is a compact subset of $(0,\infty)$.  If $\mu_q$ has density $g$, then
$\nu$ has density
\begin{equation}
h(v)=\frac{1}{v^2}g\left(-\frac{1}{v}\right)
\label{density-transform}
\end{equation}
on the corresponding interval.  Thus local real-analyticity of $g$ away from
$q=0$ is equivalent to local real-analyticity of $h$ after this change of
variables.

%%%%%%%%%%%%%%%%%%%%%%%%%%%%%%%%%%%%%%%%%%%%%%%%%%%%%%%%%%%%
\section{Effective lattice description in the strong-binding regime}\label{sec:krein}
%%%%%%%%%%%%%%%%%%%%%%%%%%%%%%%%%%%%%%%%%%%%%%%%%%%%%%%%%%%%

At negative energy, free propagation between two point scatterers is
exponentially damped with their separation.  Thus strong attraction gives both strong local binding and weak coupling
between different sites.
The Krein formula expresses the continuum problem as an effective disordered
lattice problem.

For $z\in\C\setminus[0,\infty)$ let
$$
\kappa(z)=\sqrt{-z},\qquad \Re\kappa(z)>0.
$$
The free resolvent kernel is
\begin{equation}
G_0(z;x)=\frac{1}{2\kappa(z)}e^{-\kappa(z)|x|}.
\label{free-kernel}
\end{equation}
It is convenient to use
\begin{equation}
\zeta=\varrho(z):=G_0(z;0)=\frac{1}{2\kappa(z)}.
\label{rho}
\end{equation}
Conversely, in the right half-plane,
\begin{equation}
z=z(\zeta):=-\frac{1}{4\zeta^2},
\qquad
\kappa(z(\zeta))=\frac{1}{2\zeta}.
\label{zeta-map}
\end{equation}
If $\Re\zeta>0$ and $\Im\zeta>0$, then $z(\zeta)\in\C_+$.
At the energy $z(\zeta)$ we write
\begin{equation}
g_\zeta(x):=G_0(z(\zeta);x)
 =\zeta\exp\left(-\frac{|x|}{2\zeta}\right).
\label{g-zeta}
\end{equation}

The translates $g_\zeta(\cdot-n)$ define a bounded synthesis operator
from $\ell^2(\Z)$ to $L^2(\R)$.  Define also the bounded convolution
operator $\cG(\zeta)$ on $\ell^2(\Z)$ by
$$
\cG(\zeta;n,m)=g_\zeta(n-m).
$$
We separate its diagonal and off-diagonal parts and set
\begin{equation}
T(\zeta;n,m)=
\begin{cases}
 g_\zeta(n-m),&n\ne m,\\
 0,&n=m,
\end{cases}
\label{Tdef}
\end{equation}
and
\begin{equation}
D_\omega(\zeta;n,m)=\bigl(V_\omega(n)-\zeta\bigr)\delta_{nm}.
\label{Ddef}
\end{equation}
Here $\delta_{nm}$ is the Kronecker delta.  The reduced lattice operator is
\begin{equation}
A_\omega(\zeta)=D_\omega(\zeta)-T(\zeta).
\label{Adef}
\end{equation}

The standard Krein formula for point interactions gives the following
representation.  We state only the form needed below.  Infinite uniformly
separated one-dimensional families, including the operator realization and
the resolvent construction, are treated in Sec.~4 of
Ref.~\refcite{AlbeverioEtAl1984} and Chap.~III.2 of
Ref.~\refcite{Albeverio2005}.

\begin{prop}[Krein formula]\label{prop:krein}
Let $z=z(\zeta)\in\C_+$ with $\Re\zeta>0$.  Then
$A_\omega(\zeta)$ is invertible and the resolvent kernel of $H_\omega$ is
\begin{equation}
(H_\omega-z)^{-1}(x,y)
 =g_\zeta(x-y)
 +\sum_{n,m\in\Z}g_\zeta(x-n)
 A_\omega(\zeta)^{-1}(n,m)g_\zeta(m-y).
\label{Krein-kernel}
\end{equation}
The series is understood in the operator sense.  Its kernel is locally
convergent after pairing with compactly supported $L^2$ functions.
Moreover, for a real $E<0$ and $\zeta=1/(2\sqrt{-E})$, with
$\rho(B)$ denoting the resolvent set of an operator $B$,
\begin{equation}
 E\in\rho(H_\omega)\quad\Longleftrightarrow\quad
 0\in\rho(A_\omega(\zeta)).
\label{spectral-correspondence}
\end{equation}
\end{prop}

\begin{proof}
For finitely many centers, the principal matrix for coupling constants
$q(n)$ has entries
$$
\bigl(q(n)^{-1}+g_\zeta(0)\bigr)\delta_{nm}
 +(1-\delta_{nm})g_\zeta(n-m).
$$
Since $q(n)^{-1}=-V(n)$ and $g_\zeta(0)=\zeta$, this matrix is
$-A_\omega(\zeta)$.  The usual finite-center resolvent formula therefore
has the sign and form in \eqref{Krein-kernel}.  The passage to the lattice of
uniformly separated centers, together with convergence in the operator sense,
follows from the infinite-center construction in Sec.~4 of
Ref.~\refcite{AlbeverioEtAl1984} and Chap.~III.2 of
Ref.~\refcite{Albeverio2005}.  For completeness, uniform invertibility at
nonreal energy can also be seen directly.  The Fourier multiplier of
$\cG(\varrho(z))$ is
\begin{equation}
 \widehat{\cG}_z(\theta)
 =\sum_{j\in\Z}\frac{1}{(\theta+2\pi j)^2-z},
 \qquad -\pi\le\theta\le\pi.
\label{G-Fourier}
\end{equation}
For $z\in\C_+$ its imaginary part is bounded below by
$\Im z/(\pi^2+|z|)^2>0$, using the term $j=0$.  Since
$A_\omega=V_\omega-\cG$, this yields a strictly negative imaginary part for
$A_\omega$, bounded away from zero uniformly in $\omega$.  Applying the same
bound to its adjoint proves invertibility and a locally uniform bound on
$\|A_\omega^{-1}\|$.  Since $E<0$ belongs to the resolvent set of the free
Laplacian, the same infinite-center resolvent construction states that
$H_\omega-E$ is invertible exactly when its principal operator
$A_\omega(1/(2\sqrt{-E}))$ is invertible.  This proves
\eqref{spectral-correspondence}; it is the Birman--Schwinger correspondence
for the form \eqref{form}.
\end{proof}

The off-diagonal part is small for strongly negative energies.  Put
\begin{equation}
s(\zeta):=\sum_{k\in\Z\setminus\{0\}}|g_\zeta(k)|.
\label{s-zeta}
\end{equation}
If $\Re(1/\zeta)>0$, then in fact
\begin{equation}
s(\zeta)
=
\frac{2|\zeta|\exp\{-\Re(1/(2\zeta))\}}
 {1-\exp\{-\Re(1/(2\zeta))\}}.
\label{s-bound}
\end{equation}
By Schur's test,
\begin{equation}
\|T(\zeta)\|\le s(\zeta).
\label{T-bound}
\end{equation}
For $\zeta\in\C_+$,
$$
\|D_\omega(\zeta)^{-1}\|\le\frac{1}{\Im\zeta}.
$$
Hence, if
\begin{equation}
s(\zeta)<\Im\zeta,
\label{initial-Neumann}
\end{equation}
then
\begin{equation}
A_\omega(\zeta)^{-1}
 =\sum_{k=0}^\infty
 \bigl(D_\omega(\zeta)^{-1}T(\zeta)\bigr)^k
 D_\omega(\zeta)^{-1}.
\label{Neumann}
\end{equation}
This expansion is used only as a starting identity in the upper half-plane.
We will not impose \eqref{initial-Neumann} on the real axis.

%%%%%%%%%%%%%%%%%%%%%%%%%%%%%%%%%%%%%%%%%%%%%%%%%%%%%%%%%%%%
\section{Local disorder averages in the binding-energy window}\label{sec:single-site}
%%%%%%%%%%%%%%%%%%%%%%%%%%%%%%%%%%%%%%%%%%%%%%%%%%%%%%%%%%%%

Let
$$
J=(a,b),\qquad 0<a<b,
$$
and let $\delta>0$ be so small that $a-\delta>0$.  We write
$$
J_0=(a-\delta,b+\delta),
\qquad
\Om_\delta(J)=\{\zeta\in\C:\dist(\zeta,J)<\delta\}.
$$
Here $\dist(\zeta,J)$ denotes the Euclidean distance from $\zeta$ to $J$.

\begin{ass}[Local analytic density]\label{ass:density}
The restriction of $\nu$ to $J_0$ has a density $h$.  The function $h$
extends holomorphically to $\Om_\delta(J)$ and continuously to its closure.
\end{ass}

No analyticity is assumed for the distribution outside $J_0$.  This local
form of the assumption is useful because a nonzero compactly supported
density cannot be holomorphic across both endpoints of its full support.

For each integer $\ell\ge1$ and $\zeta\in\C_+$ define
\begin{equation}
B_\ell(\zeta)=\int_\R\frac{d\nu(v)}{(v-\zeta)^\ell}.
\label{Bell}
\end{equation}

The contour deformation used in the proof of Lemma~\ref{lem:Bell} in
the Appendix is illustrated in
Figure~\ref{fig:single-site-contour}.  It is the same deformation as in
Ref.~\refcite{KaminagaKrishnaNakamura2012}, now applied to the inverse-coupling
variable.

\begin{figure}[t]
\centering
\IfFileExists{fig1_mod(3).pdf}{%
\includegraphics[width=0.88\textwidth]{fig1_mod(3).pdf}%
}{%
\includegraphics[width=0.88\textwidth]{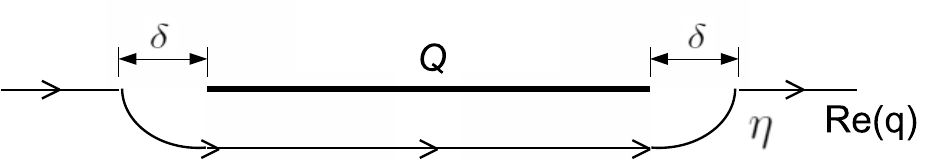}%
}
\caption{Contour deformation for the single-site average $B_\ell$.
The labels $q$ and $Q$ in the schematic correspond, respectively, to the
integration variable $v$ and the interval $J$ in the present notation.
The endpoints of the lower contour $\eta$ are $a-\delta$ and
$b+\delta$, so the real segment between them is $J_0$.}
\label{fig:single-site-contour}
\end{figure}

The continuation estimate required below is technical, so its statement and
proof are collected in Lemma~\ref{lem:Bell} of the Appendix.  In particular, $B_\ell$ extends
holomorphically from $\C_+$ to $\C_+\cup\Om_\delta(J)$ and, for every
$0<\delta_0<\delta$, satisfies the uniform bound \eqref{Bell-bound} with a
constant $C_\nu\ge1$ independent of $\ell$.

We use the same symbol $B_\ell$ for this continuation.  On the real interval
$J$, and below it, this symbol denotes the contour expression
\eqref{Bell-contour}, not the original real-line integral.  In particular,
\begin{equation}
 \Im B_1(v)=\pi h(v),\qquad v\in J,
\label{B1-boundary}
\end{equation}
by the upper-half-plane boundary value of the Cauchy transform.

In the rest of the paper we fix
$$
\delta_0=\frac{\delta}{2}
$$
and put
\begin{equation}
\Om:=\Om_{\delta/2}(J).
\label{Omega-small}
\end{equation}
Let $C_\nu$ be the constant in Lemma~\ref{lem:Bell} and set
\begin{equation}
M:=\frac{2C_\nu}{\delta}.
\label{Mdef}
\end{equation}
Since $C_\nu\ge1$, the factorization over distinct sites along a path and
\eqref{Bell-bound} imply the convenient estimate
\begin{equation}
\prod_{\alpha}
|B_{r_\alpha}(\zeta)|
\le M^{r_1+\cdots+r_s},
\label{product-B}
\end{equation}
whenever $r_1,\ldots,r_s\ge1$ and
$\zeta\in\Om$.

%%%%%%%%%%%%%%%%%%%%%%%%%%%%%%%%%%%%%%%%%%%%%%%%%%%%%%%%%%%%
\section{Weak propagation between centers and analytic DOS}\label{sec:rwe}
%%%%%%%%%%%%%%%%%%%%%%%%%%%%%%%%%%%%%%%%%%%%%%%%%%%%%%%%%%%%

A path of length $k$ from $n$ to $m$ is a sequence
$$
\gamma=(n_0,n_1,\ldots,n_k),
\qquad n_0=n,\quad n_k=m,
$$
with $n_j\ne n_{j+1}$.  Let $\Gamma_k(n,m)$ denote the set of these paths.
Jumps are not restricted to nearest neighbors,
because $T(\zeta)$ has exponentially decaying long-range matrix elements.
For $\alpha\in\Z$, let
$$
r_\alpha(\gamma)
 =\#\{j:n_j=\alpha\}.
$$
Then
\begin{equation}
\sum_{\alpha\in\Z}r_\alpha(\gamma)=k+1.
\label{visit-sum}
\end{equation}

For $\zeta$ satisfying \eqref{initial-Neumann}, the matrix element of
\eqref{Neumann} is
\begin{equation}
A_\omega(\zeta)^{-1}(n,m)
 =\sum_{k=0}^\infty\ \sum_{\gamma\in\Gamma_k(n,m)}
 \prod_{j=0}^k\frac{1}{V_\omega(n_j)-\zeta}
 \prod_{j=0}^{k-1}g_\zeta(n_j-n_{j+1}).
\label{path-expansion}
\end{equation}
The series is absolutely convergent.  Its length-$k$ absolute row sum is at
most
$$
(\Im\zeta)^{-1}
\left\{\frac{s(\zeta)}{\Im\zeta}\right\}^k.
$$
Hence expectation and the path sums may be interchanged in this initial
region.  By independence,
\begin{equation}
\E\left[
\prod_{j=0}^k\frac{1}{V_\omega(n_j)-\zeta}
\right]
 =\prod_{\alpha\in\Z}B_{r_\alpha(\gamma)}(\zeta),
\label{path-average}
\end{equation}
where factors with $r_\alpha=0$ are omitted.

Define
\begin{equation}
s_*:=\sup_{\zeta\in\Om}s(\zeta).
\label{sstar}
\end{equation}
This number is finite because $\overline\Om$ lies in the right half-plane.

\begin{ass}[Quantitative strong-attraction condition]\label{ass:small}
With $M$ and $s_*$ defined in \eqref{Mdef} and \eqref{sstar}, assume
\begin{equation}
Ms_*<1
\label{smallness-1}
\end{equation}
\end{ass}

The dimensionless product $Ms_*$ compares the size of the disorder-averaged
on-site denominators with the total propagation amplitude between sites.  The
condition $Ms_*<1$ is therefore a quantitative version of the physical
separation between strong local binding and weak propagation among different
scatterers.

This condition gives convergence after disorder averaging.  It also provides
the overlap with the elementary Neumann expansion in the upper half-plane:
since $C_\nu\ge1$ and $M=2C_\nu/\delta$, it implies
\begin{equation}
 s_*<\frac{1}{M}=\frac{\delta}{2C_\nu}
 \le\frac{\delta}{2}.
\label{overlap-consequence}
\end{equation}

The path expansion can now be continued through the real binding-energy
window after disorder averaging.  Proposition~\ref{prop:continued-inverse} in the Appendix gives a holomorphic continuation
$\cA_{nm}(\zeta)$ of
$\E[A_\omega(\zeta)^{-1}(n,m)]$ to $\Om$, uniformly in $n,m$, with the
bound \eqref{Abar-bound}.  In physical terms, the condition $Ms_*<1$ controls
the total contribution of all paths between centers after the random local
denominators have been averaged.

We now pass from the lattice inverse back to the continuum local resolvent.
For $n,m\in\Z$ define
\begin{equation}
K_{nm}(\zeta)
 =\int_0^1 g_\zeta(x-n)g_\zeta(m-x)\,d x.
\label{Knm}
\end{equation}
Each $K_{nm}$ is holomorphic in the right half-plane, by dominated
convergence on compact subsets.  Because $\overline\Om$ is a compact subset
of the right half-plane, there
are constants $R,c>0$ such that
\begin{equation}
|g_\zeta(x)|\le R e^{-c|x|},
\qquad \zeta\in\Om.
\label{g-uniform}
\end{equation}
Consequently,
\begin{equation}
\sup_{\zeta\in\Om}
\sum_{n,m\in\Z}|K_{nm}(\zeta)|<\infty.
\label{K-sum}
\end{equation}
Indeed, the left-hand side is bounded by
$$
\sup_{\zeta\in\Om}
\int_0^1
\left(\sum_{n\in\Z}|g_\zeta(x-n)|\right)^2d x.
$$

To convert the continued local resolvent into a statement about the DOS
measure, we use the standard Stieltjes-continuation result stated and proved
as Lemma~\ref{lem:stieltjes-continuation} in the Appendix.

The previous estimates quantify a regime in which local binding dominates the
residual propagation.  The next theorem shows that this separation is strong
enough to make the disorder-averaged spectral profile analytic on the
corresponding binding-energy window.

\begin{thm}[Real-analyticity of the DOS]\label{thm:main}
Assume Assumptions~\ref{ass:density} and \ref{ass:small}.  Let
\begin{equation}
I=z(J)
 =\left(-\frac{1}{4a^2},-\frac{1}{4b^2}\right).
\label{energy-I}
\end{equation}
Then the upper-half-plane Stieltjes transform $m_+(z)$ in \eqref{mplus}
admits a holomorphic continuation through $I$.  In particular, the density
of states $n(E)$ exists and is real-analytic on $I$.  The IDS $N(E)$ is also
real-analytic on $I$.
\end{thm}

\begin{proof}
For $\zeta$ in the open subset of $\Om\cap\C_+$ on which the Neumann
expansion is valid, Proposition~\ref{prop:krein} gives
\begin{equation}
\begin{split}
m_+(z(\zeta))
={}&\zeta\\
&+\sum_{n,m\in\Z}K_{nm}(\zeta)
\E\bigl[A_\omega(\zeta)^{-1}(n,m)\bigr].
\end{split}
\label{trace-series}
\end{equation}
Here the free contribution is $\int_0^1g_\zeta(0)\,d x=\zeta$.
All traces, expectations, and sums in \eqref{trace-series} may be
interchanged because the initial Neumann series and the two sums involving
$g_\zeta$ are absolutely convergent.  By
Proposition~\ref{prop:continued-inverse}, \eqref{Abar-bound}, and
\eqref{K-sum}, the right-hand side of
\begin{equation}
\widetilde m(\zeta)
 =\zeta+\sum_{n,m\in\Z}K_{nm}(\zeta)\cA_{nm}(\zeta)
\label{mtilde}
\end{equation}
converges absolutely and uniformly on compact subsets of $\Om$ and defines
a holomorphic function there.  On the above open subset of
$\Om\cap\C_+$, it agrees with $m_+(z(\zeta))$.  Both functions are
holomorphic on the connected set $\Om\cap\C_+$.  The identity theorem shows
that they agree throughout that set.  Thus $\widetilde m$ is the analytic
continuation of $m_+(z(\zeta))$ in the $\zeta$ variable.

The map $z(\zeta)=-1/(4\zeta^2)$ is biholomorphic on a sufficiently small
right-half-plane neighborhood of $J$.  Therefore $m_+$ has a holomorphic
continuation to a complex neighborhood of the real interval $I=z(J)$.
For $E\in I$, put $\zeta=\varrho(E)=1/(2\sqrt{-E})\in J$.  The resulting
continuation in the energy variable is
$$
m^{\mathrm{ext}}(z)=\widetilde m(\varrho(z))
$$
near $I$.  Lemma~\ref{lem:stieltjes-continuation} gives
\begin{equation}
n(E)=\frac{1}{\pi}\Im\widetilde m(\varrho(E)).
\label{analytic-density}
\end{equation}
The right-hand side is real-analytic in $E$.  Moreover, for any fixed
$E_0\in I$,
$$
 N(E)=N(E_0)+\int_{E_0}^E n(t)\,d t,
$$
so $N$ is real-analytic on $I$ as well.
\end{proof}

\begin{rem}
Theorem~\ref{thm:main} does not assume a pointwise lower bound on
$|V_\omega(n)-\zeta|$ for real $\zeta\in J$.  This is essential.  Such a
bound would remove the relevant values of the random diagonal variable and
could put the corresponding continuum energy in a gap.  The denominator is
moved away from the real singularity only inside the disorder integral, by
the contour deformation in Lemma~\ref{lem:Bell}.
\end{rem}

%%%%%%%%%%%%%%%%%%%%%%%%%%%%%%%%%%%%%%%%%%%%%%%%%%%%%%%%%%%%
\section{Spectral location and controlled atomic limit}\label{sec:spectrum}
%%%%%%%%%%%%%%%%%%%%%%%%%%%%%%%%%%%%%%%%%%%%%%%%%%%%%%%%%%%%

The map from inverse coupling to energy has a direct one-center
interpretation.  Since $V=-1/q$, Eq.~\eqref{single-bound-state} gives
\begin{equation}
E_{\rm b}(q)=-\frac{q^2}{4}
=-\frac{1}{4V^2}=z(V).
\label{binding-map}
\end{equation}
Thus the interval $I=z(J)$ is the image, under the isolated-scatterer
binding-energy relation, of the inverse-coupling interval $J$.  The support condition used below ensures that this entire window belongs
to the almost-sure spectrum.  A quantitative statement about its DOS is
provided separately by the atomic-limit approximation below.

The preceding theorem is most useful when $I$ is contained in the spectrum,
rather than in a resolvent interval.  In the present model this has a simple
interpretation.

\begin{prop}[A nontrivial spectral interval]\label{prop:spectrum}
Assume, in addition to the hypotheses of Theorem~\ref{thm:main}, that
\begin{equation}
J\subset\operatorname{int}(\supp\nu).
\label{J-support}
\end{equation}
Here $\operatorname{int}$ denotes interior and $\sigma(B)$ denotes the spectrum
of an operator $B$.  Then
$$
I=z(J)\subset\sigma(H_\omega)
$$
almost surely.
\end{prop}

\begin{proof}
We construct a Weyl sequence for the reduced operator.  This argument does
not require a pointwise inverse on the real axis.  Fix first
$v\in J$ and put
$$
r=e^{-1/(2v)}\in(0,1).
$$
At the real parameter $v$, the convolution operator $T(v)$ has Fourier
multiplier
\begin{equation}
\tau_v(\theta)
 =2v\sum_{k=1}^{\infty}r^k\cos(k\theta)
 =2v\frac{r\cos\theta-r^2}
 {1-2r\cos\theta+r^2}.
\label{T-symbol}
\end{equation}
Thus $\tau_v(\theta_v)=0$ for
$\theta_v=\arccos r$.  Define
\begin{equation}
 c_L(n)=\frac{1}{\sqrt{2L+1}}e^{in\theta_v}
 \boldsymbol 1_{[-L,L]}(n),
\label{cutoff-wave}
\end{equation}
where $\boldsymbol 1_{[-L,L]}(n)$ is the indicator of the integer interval
$\{n\in\Z:|n|\le L\}$.  We have
\begin{equation}
 \|T(v)c_L\|\longrightarrow0
 \quad\hbox{as }L\longrightarrow\infty.
\label{T-Weyl}
\end{equation}
Indeed, the convolution kernel of $T(v)$ belongs to $\ell^1(\Z)$.
To verify \eqref{T-Weyl}, truncate this kernel to a range $R$.  Away from a
boundary of width $R$, its action on \eqref{cutoff-wave} is multiplication by
the truncated Fourier multiplier.  Since the full multiplier vanishes at
$\theta_v$, the absolute value of the truncated multiplier is bounded by the
$\ell^1$ norm of the omitted tail.  The relative size of the boundary tends
to zero as $L\to\infty$.  Letting first $L\to\infty$ and then $R\to\infty$
proves \eqref{T-Weyl}.

Since $v\in\supp\nu$, for every $\epsilon>0$ and every $L$ a prescribed
block of $2L+1$ variables has a positive probability of satisfying
\begin{equation}
 \max_{|n-k|\le L}|V_\omega(n)-v|<\epsilon.
\label{good-block}
\end{equation}
Fix sequences $L_j\to\infty$ and $\epsilon_j\downarrow0$.  For each $j$,
consider infinitely many mutually disjoint blocks of length $2L_j+1$.
The corresponding events \eqref{good-block} are independent and all have the
same positive probability.  The Borel-Cantelli lemma shows that infinitely
many of them occur almost surely.  Taking the intersection of these
probability-one events over $j$, we may choose one successful block for each
$j$, with centers $k_j$ tending to infinity.  Translate
$c_{L_j}$ to the block centered at $k_j$ and denote the resulting unit vector
by $c_j$.  Translation invariance of $T(v)$ and \eqref{good-block} give
\begin{equation}
 \|A_\omega(v)c_j\|
 \le \epsilon_j+\|T(v)c_{L_j}\|\longrightarrow0.
\label{A-Weyl}
\end{equation}
Hence $0\in\sigma(A_\omega(v))$.  By
\eqref{spectral-correspondence}, $z(v)\in\sigma(H_\omega)$.

To remove the dependence of the null set on $v$, apply the preceding argument
to a countable dense subset $J_{\mathrm d}$ of $J$ and intersect the resulting
probability-one events.  Almost surely, the closed set $\sigma(H_\omega)$
contains $z(J_{\mathrm d})$.  Since $z$ is continuous on $J$,
it also contains its closure and therefore contains $I=z(J)$.
\end{proof}

The smallness condition in Assumption~\ref{ass:small} is naturally a
strong-attraction condition.  This can be seen directly from the bound
\eqref{s-bound}.  For real $\zeta>0$,
\begin{equation}
s(\zeta)
 =\frac{2\zeta e^{-1/(2\zeta)}}{1-e^{-1/(2\zeta)}}.
\label{s-real}
\end{equation}
Thus $s(\zeta)$ is exponentially small as $\zeta\downarrow0$.  Since
$q=-1/V$, small inverse coupling $V$ corresponds to large attractive
coupling $|q|$.  Under the scaling $q\mapsto\lambda q$, the inverse coupling
is of order $\lambda^{-1}$, while the single-center binding energy is of order
$\lambda^2$.  At the same time, the hopping factor
$g_\zeta(k)=\zeta\exp(-|k|/(2\zeta))$ is exponentially small in $\lambda$ for
$k\ne0$.  This is the physical origin of the scale separation used below.

\begin{cor}[Strong-attraction scaling]\label{cor:scaling}
Let $\widetilde q_\omega(n)$ be i.i.d. random variables supported in a
compact interval of $(-\infty,0)$, and suppose the density of
$\widetilde V=-1/\widetilde q$ is real-analytic on an open neighborhood of
the closure of an interval
$\widetilde J=(\widetilde a,\widetilde b)\subset(0,\infty)$.  Define
$$
q_\omega^{(\lambda)}(n)=\lambda\widetilde q_\omega(n),
\qquad \lambda>0.
$$
Then, for all sufficiently large $\lambda$, Theorem~\ref{thm:main} applies
to an interval
$$
J_\lambda=\lambda^{-1}\widetilde J
$$
and hence to the negative-energy interval
$$
I_\lambda
 =\left\{-\frac{1}{4\zeta^2}:\zeta\in J_\lambda\right\}.
$$
If $\overline{\widetilde J}$ is contained in the interior of the support of
$\widetilde V$, then $I_\lambda$ is contained in the almost-sure spectrum.
\end{cor}

\begin{proof}
The inverse coupling scales as
$$
V_\omega^{(\lambda)}=\lambda^{-1}\widetilde V_\omega.
$$
Let $\widetilde h$ denote the density of $\widetilde V$.  By compactness of
$\overline{\widetilde J}$ and real-analyticity, one can choose
$0<\widetilde\delta<\widetilde a$ such that $\widetilde h$ has a holomorphic extension to
$\Om_{\widetilde\delta}(\widetilde J)$ which is continuous on its closure.
For the scaled law the density and the analytic width are
\begin{equation}
 h_\lambda(v)=\lambda\widetilde h(\lambda v),
 \qquad \delta_\lambda=\lambda^{-1}\widetilde\delta.
\label{scaled-density}
\end{equation}
In the constant in Lemma~\ref{lem:Bell}, the length of the scaled contour is
of order $\lambda^{-1}$ and the supremum of $|h_\lambda|$ is of order
$\lambda$.  The constants $C_{\nu_\lambda}$ may therefore be chosen uniformly
in $\lambda$, and
\begin{equation}
 M_\lambda=\frac{2C_{\nu_\lambda}}{\delta_\lambda}
 =O(\lambda).
\label{scaled-M}
\end{equation}
Let $s_{*,\lambda}$ denote the quantity in \eqref{sstar} for the scaled
neighborhood.
On the other hand, write $\zeta=w/\lambda$ on the scaled complex
neighborhood.  Its unscaled closure is a compact subset of the right
half-plane, so $\Re(1/w)$ has a positive lower bound.  Formula
\eqref{s-bound} gives, uniformly on this neighborhood,
$$
s(\zeta)\le C\lambda^{-1}e^{-c\lambda}
$$
with some $C,c>0$.  Hence $M_\lambda s_{*,\lambda}\to0$ as
$\lambda\to\infty$.  Assumption~\ref{ass:small}, and therefore the hypotheses
of Theorem~\ref{thm:main}, hold for sufficiently large $\lambda$.
The spectral statement follows from Proposition~\ref{prop:spectrum}.
\end{proof}

Physically, the next result quantifies the approach to an ensemble of nearly
isolated bound states as the attraction strength increases.  It gives an error bound for the atomic approximation.

\begin{prop}[Atomic limit with an exponential error]\label{prop:atomic}
Use the scaled laws of Corollary~\ref{cor:scaling}, and write $n_\lambda$ for
the corresponding DOS.  Put
$$
 \widetilde I=z(\widetilde J),\qquad
 w(\xi)=\frac{1}{2\sqrt{-\xi}}.
$$
For every compact interval $K\subset\widetilde I$, there are constants
$C_K,c_K>0$ such that, for all sufficiently large $\lambda$,
\begin{equation}
 \sup_{\xi\in K}
 \left|\lambda^2 n_\lambda(\lambda^2\xi)
       -2w(\xi)^3\widetilde h(w(\xi))\right|
 \le C_K e^{-c_K\lambda}.
\label{atomic-error}
\end{equation}
The leading term is the density of the isolated-center energy
$-\widetilde q^2/4$.  If $\widetilde q$ has density $\widetilde g$ near the
corresponding couplings, it equals
\begin{equation}
 n_{\rm at}(\xi)
 =2w(\xi)^3\widetilde h(w(\xi))
 =\frac{\widetilde g(-2\sqrt{-\xi})}{\sqrt{-\xi}}.
\label{atomic-density}
\end{equation}
In particular, if $\widetilde h$ is strictly positive on $w(K)$, then
$n_\lambda(\lambda^2\xi)>0$ on $K$ for all sufficiently large $\lambda$.
\end{prop}

\begin{proof}
The path of length zero contributes
$B_1(\zeta)\delta_{nm}$ to $\cA_{nm}$.  The remaining paths satisfy
\begin{equation}
 \left|\cA_{nm}(\zeta)-B_1(\zeta)\delta_{nm}\right|
 \le\frac{M^2s_*}{1-Ms_*},\qquad\zeta\in\Om,
\label{A-remainder}
\end{equation}
by the geometric majorant in Proposition~\ref{prop:continued-inverse}.
Also, by translation and \eqref{g-zeta},
\begin{equation}
 \sum_{n\in\Z}K_{nn}(\zeta)
 =\int_\R g_\zeta(x)^2\,d x=2\zeta^3.
\label{K-diagonal-sum}
\end{equation}
Thus the zero-length part of \eqref{mtilde} is
$\zeta+2\zeta^3B_1(\zeta)$.  For real $\zeta>0$ the kernels in
\eqref{Knm} are nonnegative.  Grouping their sum by $j=m-n$ gives
\begin{equation}
\begin{split}
 S(\zeta):=\sum_{n,m\in\Z}K_{nm}(\zeta)
 &=\sum_{j\in\Z}\int_\R g_\zeta(x)g_\zeta(x-j)\,d x\\
 &=2\zeta^3\frac{1+r}{1-r}
   +2\zeta^2\frac{r}{(1-r)^2},\qquad
 r=e^{-1/(2\zeta)}.
\end{split}
\label{K-total-sum}
\end{equation}
Here the convolution integral is
$\zeta^2(|j|+2\zeta)e^{-|j|/(2\zeta)}$.
For $\zeta=w/\lambda$ with $w$ in a compact positive interval,
$S(\zeta)\le C\lambda^{-3}$ for large $\lambda$.
Corollary~\ref{cor:scaling} gives
$M_\lambda\le C\lambda$ and
$s_{*,\lambda}\le C\lambda^{-1}e^{-c\lambda}$.
Combining \eqref{A-remainder} and \eqref{K-total-sum}, we obtain
\begin{equation}
 \left|\widetilde m_\lambda(\zeta)
       -\zeta-2\zeta^3B_{1,\lambda}(\zeta)\right|
 \le C\lambda^{-2}e^{-c\lambda}
\label{m-atomic-remainder}
\end{equation}
on the corresponding real interval.  Taking imaginary parts, using
\eqref{B1-boundary} and $h_\lambda(w/\lambda)=\lambda\widetilde h(w)$,
and multiplying by $\lambda^2$ proves \eqref{atomic-error}.
Equation~\eqref{atomic-density} follows from \eqref{density-transform}.
Positivity follows from the positive minimum of the leading term on $K$.
\end{proof}

The energy change in \eqref{atomic-density} includes the Jacobian
$d w/d\xi=2w^3$.  Thus a unit mass of isolated bound states gives one state
per lattice cell.  The approximation describes how a distribution of local
binding energies forms the leading negative-energy DOS, while propagation
between centers changes this profile by an exponentially small amount on the
rescaled interior window.  It also specifies the role of the regularity
assumption: local analyticity of the binding-energy
distribution remains valid after coupling between the centers is included.  No
comparable smoothing assertion is made for arbitrary nonanalytic laws.

\begin{rem}[Uniform coupling law]\label{rem:uniform-coupling}
If $q$ is uniformly distributed on $[q_-,q_+]\subset(-\infty,0)$, then
$V=-1/q$ has density
$$
h(v)=\frac{1}{(q_+-q_-)v^2}
$$
on
$$
\left[-\frac{1}{q_-},-\frac{1}{q_+}\right].
$$
The function $v\mapsto v^{-2}$ is holomorphic away from zero.  Therefore
Assumption~\ref{ass:density} holds for every open interval $J$ whose closure
is contained strictly inside this inverse-coupling interval.  After the strong-attraction
scaling of Corollary~\ref{cor:scaling}, Assumption~\ref{ass:small} is also
satisfied for sufficiently large coupling strength.
\end{rem}

%%%%%%%%%%%%%%%%%%%%%%%%%%%%%%%%%%%%%%%%%%%%%%%%%%%%%%%%%%%%
\section{Numerical strong-coupling regime and DOS profile}\label{sec:numerics}
%%%%%%%%%%%%%%%%%%%%%%%%%%%%%%%%%%%%%%%%%%%%%%%%%%%%%%%%%%%%

The analytic result gives a sufficient strong-attraction condition, but the
size of this regime and the position of the corresponding energy window are
best understood on the physical binding-energy scale.  We therefore examine a
uniform coupling distribution for which both quantities can be displayed
explicitly.  First we estimate the coupling strength at which
Assumption~\ref{ass:small} is rigorously guaranteed.  We then compare the
resulting interval with a finite-volume DOS computed directly from the exact
point-interaction recurrence.  The numerical calculation is used only to show
where the interval covered by the theorem is located in the spectral profile; it is
not used in the proof.

Let
$$
\widetilde q_\omega(n)\sim {\rm Unif}[-2,-1],\qquad
q_\omega^{(\lambda)}(n)=\lambda\widetilde q_\omega(n).
$$
Then $\widetilde V=-1/\widetilde q$ is supported on $[1/2,1]$ and has density
$$
\widetilde h(v)=\frac{1}{v^2},\qquad \frac12\le v\le1.
$$
We choose
$$
\widetilde J=(0.65,0.85),\qquad
\widetilde\delta=0.10.
$$
Thus
$$
J_\lambda=\lambda^{-1}\widetilde J,\qquad
\delta_\lambda=\lambda^{-1}\widetilde\delta.
$$
For the contour in Lemma~\ref{lem:Bell},
$$
\operatorname{length}(\widetilde\eta)=0.20+0.10\pi,\qquad
\sup_{w\in\widetilde\eta}|\widetilde h(w)|\le \frac{1}{0.55^2}.
$$
Hence the constant in Lemma~\ref{lem:Bell} may be bounded by
$$
C_{\nu_\lambda}\le 1+\frac{0.20+0.10\pi}{0.55^2}=2.699700\ldots,
$$
uniformly in $\lambda$.  Therefore
\begin{equation}
M_\lambda\le 53.994001\ldots\,\lambda.
\label{numerical-M-bound}
\end{equation}
On the scaled neighborhood write $\zeta=w/\lambda$.  For $w\in\overline{\Om_{0.05}(\widetilde J)}$, we have $|w|\le0.90$.  To see the required reciprocal bound, write
$w=u+it$ and choose $x\in[0.65,0.85]$ with $|w-x|\le0.05$.
Then
$$
 |w|^2-(x+0.05)u
 \le (x-0.05)(u-x-0.05)\le0.
$$
Since $u>0$ and $x+0.05\le0.90$, this proves
$$
\Re\frac{1}{w}\ge\frac{1}{0.90}=\frac{10}{9},
$$
with equality at $w=0.90$.  Hence
$$
\Re\frac{1}{2\zeta}
=\lambda\Re\frac{1}{2w}
\ge\frac{5}{9}\lambda.
$$
Using \eqref{s-bound}, we obtain
\begin{equation}
s_{*,\lambda}\le \frac{1.8}{\lambda}
\frac{e^{-5\lambda/9}}{1-e^{-5\lambda/9}}.
\label{numerical-s-bound}
\end{equation}
Combining this with \eqref{numerical-M-bound} gives
\begin{equation}
M_\lambda s_{*,\lambda}\le 97.189202\ldots
\frac{e^{-5\lambda/9}}{1-e^{-5\lambda/9}}.
\label{numerical-smallness-bound}
\end{equation}
The right-hand side of \eqref{numerical-smallness-bound} can be written as
$$
\frac{97.189202\ldots}{e^{5\lambda/9}-1}.
$$
It is strictly decreasing in $\lambda$ and is less than one whenever
$$
\lambda>\frac{9}{5}\log(98.189202\ldots)=8.2564\ldots .
$$
Thus this explicit, and rather rough, sufficient estimate verifies Assumption~\ref{ass:small} for every integer $\lambda\ge9$.  This does not imply that the condition fails for smaller values of $\lambda$.

We next give a finite-volume numerical illustration without replacing the point interactions by regular potentials.  Consider $(0,N+1)$ with Dirichlet boundary conditions and point interactions at $1,\ldots,N$.  For $E=-\kappa^2<0$, the free equation and the jump condition \eqref{jump} give the exact recurrence
\begin{equation}
u_{n+1}+u_{n-1}
=\left(
2\cosh\kappa
+\frac{q_\omega^{(\lambda)}(n)}{\kappa}\sinh\kappa
\right)u_n.
\label{numerical-recurrence}
\end{equation}
Starting from $u_0=0$ and $u_1=1$, the number of eigenvalues strictly below
$E$ is the number of zeros of the corresponding solution in $(0,N+1)$,
by Sturm oscillation.  For $E<0$, on each free interval,
\begin{equation}
 u(n+t)=\frac{\sinh(\kappa(1-t))}{\sinh\kappa}u_n
       +\frac{\sinh(\kappa t)}{\sinh\kappa}u_{n+1},
 \qquad 0\le t\le1.
\label{interval-interpolation}
\end{equation}
The two coefficients are positive in the interior.  Thus, away from exact
zeros at a site, each sign change between successive values gives exactly
one zero and all other intervals give none.  Exact zeros at a site are handled by a one-sided limit in $E$.
Taking this limit from above gives the count of eigenvalues less than or
equal to $E$, as required for the right-continuous IDS.  A common positive rescaling of successive values
can prevent overflow without changing the count.  Thus no regular-potential
approximation of the delta interactions is needed.  The count divided by
$N+1$, the physical length, defines the finite-volume IDS per unit length;
normalization per scatterer instead differs by the factor $(N+1)/N$.

We take $N=200$ and average over $300$ independent disorder realizations.  The
same $300$ samples of the unscaled variables $\widetilde q_\omega(n)$ are used
for all four values of $\lambda$.  This common-random-number choice does not
change any marginal estimator, but it reduces sampling noise when the curves
for different values of $\lambda$ are compared.  The scaled energy
$$
\xi=\frac{E}{\lambda^2}
$$
is sampled at $900$ equally spaced points in $[-0.72,-0.25]$.  The resulting finite-volume IDS is averaged over the disorder.  For the DOS
plot, this averaged empirical IDS is smoothed with a Gaussian of standard
deviation $\sigma_\xi=0.004$ and then differentiated numerically.  In the
recurrence, the coefficient is evaluated in the equivalent form
$$
 \left(1+\frac{q_\omega^{(\lambda)}(n)}{2\kappa}\right)e^\kappa
 +\left(1-\frac{q_\omega^{(\lambda)}(n)}{2\kappa}\right)e^{-\kappa},
$$
which avoids loss of relative accuracy near the isolated-center condition
$q_\omega^{(\lambda)}(n)=-2\kappa$.  Successive pairs of solution values are
rescaled by a common positive factor, and sign changes are detected directly
from their signs.  Thus the rescaling cannot affect the eigenvalue count.
The figure displays a smoothed finite-sample estimator of the scaled DOS, whose
infinite-volume target, where the DOS exists, is
\begin{equation}
 \frac{d}{d\xi}N_\lambda(\lambda^2\xi)
 =\lambda^2 n_\lambda(\lambda^2\xi).
\label{scaled-DOS-target}
\end{equation}
The displayed estimator is not equal to this target at finite volume, finite
sample size, and nonzero smoothing width.  Its smooth appearance cannot
establish real-analyticity.  Smoothing and differentiation near the ends of
the sampled energy interval depend on the numerical boundary treatment.
These end regions are outside the interval used for the analytic comparison.

For the present choice of $\widetilde J$, the interval in scaled energy is independent of $\lambda$:
\begin{equation}
\frac{I_\lambda}{\lambda^2}
=\left(
-\frac{1}{4(0.65)^2},
-\frac{1}{4(0.85)^2}
\right)
=(-0.591716\ldots,-0.346021\ldots).
\label{scaled-analyticity-interval}
\end{equation}

\begin{figure}[ht]
\centering
\includegraphics[width=0.82\textwidth]{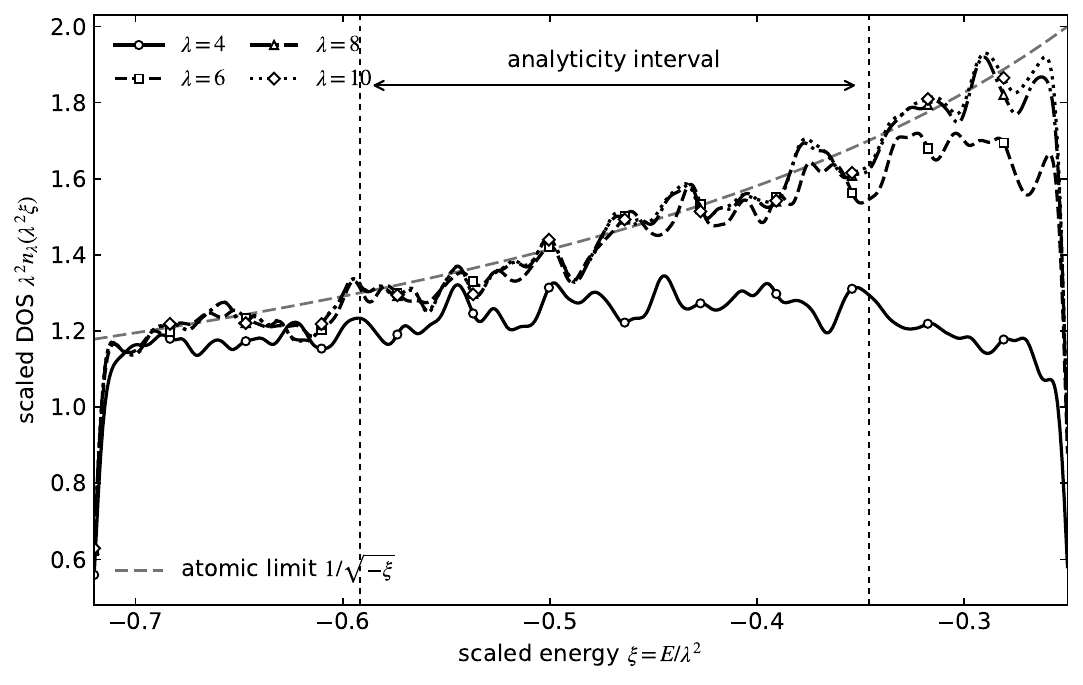}
\caption{Smoothed finite-volume scaled DOS for $\lambda=4,6,8,10$,
computed with $N=200$ point interactions and $300$ independent disorder
realizations.  The same samples of $\widetilde q_\omega(n)$ are used for all
four values of $\lambda$.  The averaged finite-volume IDS is smoothed with
$\sigma_\xi=0.004$ before differentiation.  The gray dashed curve is the
atomic-limit density $1/\sqrt{-\xi}$, and the vertical dashed lines delimit
the interval in \eqref{scaled-analyticity-interval}.  The rapid downturns at
the ends of the sampled interval result from the finite sampled energy range
and are not used to identify spectral edges.}
\label{fig:dos-num}
\end{figure}

Figure~\ref{fig:dos-num} shows the smoothed finite-volume DOS for
$\lambda=4,6,8,10$.  The two vertical dashed lines indicate the interval in
\eqref{scaled-analyticity-interval}, and the gray dashed curve shows the atomic
limit in \eqref{uniform-atomic-DOS}.  The curves for $\lambda=4,6,8$ are
included for comparison; among the values shown, the explicit bound above
verifies the hypotheses of Theorem~\ref{thm:main} for $\lambda=10$.  The
interval lies in a region with visible spectral weight, consistently with
Proposition~\ref{prop:spectrum}.  The approach of the large-$\lambda$ curves
to the atomic profile also illustrates Proposition~\ref{prop:atomic}.  The
figure is intended only as a numerical illustration of these spectral
features, not as a numerical verification of real-analyticity or of the
exponential error rate.
For this uniform law, Proposition~\ref{prop:atomic} gives the explicit
interior limiting profile
\begin{equation}
 n_{\rm at}(\xi)=\frac{1}{\sqrt{-\xi}},\qquad -1<\xi<-\frac14.
\label{uniform-atomic-DOS}
\end{equation}
The corresponding isolated-center IDS is $2-2\sqrt{-\xi}$ on this interval.
The error bound \eqref{atomic-error} is asserted on compact subintervals of
\eqref{scaled-analyticity-interval}, where the analytic hypotheses have been
verified for large $\lambda$.  At the two ends of the indicated analytic
window, the atomic DOS has limiting values $1.30$ and $1.70$.  These values
provide an analytic scale for the spectral weight in the numerical curves.
The rigorous threshold $\lambda\ge9$ ensures analyticity; it does not by
itself specify a small numerical relative error in the atomic approximation.

%%%%%%%%%%%%%%%%%%%%%%%%%%%%%%%%%%%%%%%%%%%%%%%%%%%%%%%%%%%%
\section{Discussion and conclusion}\label{sec:discussion}
%%%%%%%%%%%%%%%%%%%%%%%%%%%%%%%%%%%%%%%%%%%%%%%%%%%%%%%%%%%%

We have studied the negative-energy DOS of a one-dimensional random
Kronig-Penney model in a regime in which each attractive center is strongly
binding and propagation between distinct centers is exponentially weak at the
relevant energies.  The continuum nature of the model is retained throughout:
the delta interactions are not replaced by regular potentials, and the
finite-volume calculation uses the exact point-interaction recurrence.

The main physical conclusion is a controlled atomic limit for the DOS in this
singular continuum system, together with stability of its local analyticity
under weak propagation between centers.  The Krein
reduction separates two effects.  The random inverse couplings form the
Anderson-type diagonal part, while free propagation between different lattice
sites generates exponentially decaying hopping.  When the attraction is
strong, the hopping is sufficiently small that the disorder-averaged
random-walk expansion can be analytically continued through a real
negative-energy interval.  The resulting DOS and IDS are real-analytic there.  In addition,
Proposition~\ref{prop:atomic} identifies the leading scaled DOS as the
probability density of isolated binding energies and bounds its error
exponentially in the attraction strength.  This makes the spectral profile
quantitative, beyond its regularity alone.

The spectral inclusion result is important for the interpretation.  Under the
support condition in Proposition~\ref{prop:spectrum}, the analytic interval is
contained in the almost-sure spectrum.  Hence the result is not caused by the
absence of states.  In the uniform example of Sec.~\ref{sec:numerics}, the
explicit sufficient estimate gives $\lambda\ge9$, and the corresponding
scaled interval lies between approximately $-0.592$ and $-0.346$.  The
finite-volume data show that this window carries visible spectral weight.  The
threshold is only sufficient and is expected to be conservative.

The present result concerns spectral regularity, not localization or
transport.  No conclusion about extended states, conductivity, or a
localization length follows from real-analyticity of the DOS alone.  Further questions are how the analyticity interval is related to
transport properties, and whether similar regularity results hold for
correlated couplings, random point positions, or less restrictive
single-site distributions.  These questions require methods beyond the
disorder-averaged continuation argument used here.

\section*{Data Availability Statement}
The Python program used for the exact recurrence calculation 
and the numerical data underlying Fig. 2 are provided as Supplementary Material. 
The random seed and all numerical parameters are specified in the program, 
so that the calculation is reproducible.

%%%%%%%%%%%%%%%%%%%%%%%%%%%%%%%%%%%%%%%%%%%%%%%%%%%%%%%%%%%%
\appendix
\section{Technical continuation estimates}\label{app:continuation}
%%%%%%%%%%%%%%%%%%%%%%%%%%%%%%%%%%%%%%%%%%%%%%%%%%%%%%%%%%%%

This appendix gives the analytic continuation estimates used in the main
text.

\begin{lem}[Single-site continuation]\label{lem:Bell}
Assume Assumption~\ref{ass:density}.  Then $B_\ell$ admits a holomorphic
continuation from $\C_+$ to $\C_+\cup\Om_\delta(J)$.  Moreover, there is a constant $C_\nu\ge1$, independent of
$\ell$ and $\delta_0$, such that, for every $0<\delta_0<\delta$,
\begin{equation}
|B_\ell(\zeta)|
\le C_\nu(\delta-\delta_0)^{-\ell},
\qquad
\zeta\in\Om_{\delta_0}(J),\quad \ell\ge1.
\label{Bell-bound}
\end{equation}
\end{lem}

\begin{proof}
Let $\eta$ be the part of the boundary of the closed stadium
$\overline{\Om_\delta(J)}$ lying in the closed lower half-plane, including
its two real endpoints, and orient it from $a-\delta$ to $b+\delta$.
For $\zeta\in\C_+$, split the integral into $\R\setminus J_0$ and $J_0$.
The pole $w=\zeta$ is not in the region between $J_0$ and $\eta$.
Thus, as indicated in Figure~\ref{fig:single-site-contour}, Cauchy's theorem,
first on contours lying inside $\Om_\delta(J)$ and then by passage to the
boundary, gives
\begin{equation}
B_\ell(\zeta)
 =\int_{\R\setminus J_0}\frac{d\nu(v)}{(v-\zeta)^\ell}
  +\int_\eta\frac{h(w)\,d w}{(w-\zeta)^\ell}.
\label{Bell-contour}
\end{equation}
The two sets $\R\setminus J_0$ and $\eta$ are disjoint from
$\C_+\cup\Om_\delta(J)$.  Thus the right-hand side is holomorphic on this
connected open set and is the required continuation.

If $\zeta\in\Om_{\delta_0}(J)$, then both
$\R\setminus J_0$ and $\eta$ have distance at least
$\delta-\delta_0$ from $\zeta$.  Therefore
$$
|B_\ell(\zeta)|
\le
\left(1+\ellength(\eta)\sup_{w\in\eta}|h(w)|\right)
(\delta-\delta_0)^{-\ell}.
$$
Thus we may choose
\begin{equation}
 C_\nu=1+(b-a+\pi\delta)\sup_{w\in\eta}|h(w)|.
\label{Cnu-explicit}
\end{equation}
This constant is independent of $\ell$ and $\delta_0$, and proves
\eqref{Bell-bound}.
\end{proof}

\begin{prop}[Continued averaged inverse]\label{prop:continued-inverse}
Under Assumptions~\ref{ass:density} and \ref{ass:small}, for every
$n,m\in\Z$ the function
$$
\E\bigl[A_\omega(\zeta)^{-1}(n,m)\bigr],
$$
initially considered in the region where \eqref{initial-Neumann} holds,
has a holomorphic continuation $\cA_{nm}(\zeta)$ to $\Om$.  The continued
function is given by the path series obtained from \eqref{path-expansion},
with the random vertex product replaced by the right-hand side of
\eqref{path-average}.  Moreover,
\begin{equation}
|\cA_{nm}(\zeta)|
\le\frac{M}{1-Ms_*},
\qquad \zeta\in\Om,
\label{Abar-bound}
\end{equation}
uniformly in $n,m$.
\end{prop}

\begin{proof}
For a path $\gamma$ of length $k$, define its continued vertex factor by
\begin{equation}
 F_\gamma(\zeta)
 :=\prod_{\alpha:r_\alpha(\gamma)>0}
 B_{r_\alpha(\gamma)}(\zeta).
\label{continued-vertex}
\end{equation}
This agrees with the expectation in \eqref{path-average} in the upper
half-plane.  On all of $\Om$, \eqref{product-B} and \eqref{visit-sum} give
\begin{equation}
 |F_\gamma(\zeta)|\le M^{k+1}.
\label{vertex-bound}
\end{equation}
The estimate concerns the continued average; it is not an estimate of the
expectation of the absolute value of the random denominators on the real axis.
The sum of the absolute values of all edge weights of length $k$, with fixed
initial site and with the terminal site either fixed or summed over, is
bounded by $s_*^k$.  This follows either by
iterated convolution or by the row-sum bound for the nonnegative kernel
$|T(\zeta;n,m)|$.  Hence the absolute value of the $k$-th averaged path
contribution is bounded by
$$
M^{k+1}s_*^k.
$$
The series is therefore absolutely and uniformly convergent on $\Om$ by
\eqref{smallness-1}.  Every term is holomorphic by
Lemma~\ref{lem:Bell} and \eqref{g-zeta}, so the sum is holomorphic.
Summing the geometric majorant gives \eqref{Abar-bound}.

It remains to identify this holomorphic function as a continuation of the
actual averaged inverse.  By \eqref{overlap-consequence}, choose $x\in J$
and $y$ such that
$$
s_*<y<\frac{\delta}{2}.
$$
Then $\zeta=x+iy\in\Om\cap\C_+$ and
$$
\|D_\omega(\zeta)^{-1}T(\zeta)\|
\le\frac{s(\zeta)}{y}
\le\frac{s_*}{y}<1.
$$
Thus the Neumann series \eqref{Neumann} is valid in a neighborhood of this
point contained in $\Om\cap\C_+$, and there the continued path series equals
the actual expectation.  The uniform inverse bound in the proof of Proposition~\ref{prop:krein}
shows that the actual expectation is holomorphic throughout
$\Om\cap\C_+$.  The identity theorem therefore gives equality on that
connected set and identifies the continuation uniquely.
\end{proof}

\begin{lem}[Stieltjes continuation]\label{lem:stieltjes-continuation}
Let $U\subset\R$ be an open interval.  Suppose that the function $m_+$ in
\eqref{Stieltjes} has a holomorphic continuation $m^{\mathrm{ext}}$ to a
complex neighborhood of $U$.  Then
\begin{equation}
 d n(E)=\frac{1}{\pi}\Im m^{\mathrm{ext}}(E)\,d E
 \quad\hbox{on }U.
\label{measure-density}
\end{equation}
In particular, the density on the right-hand side is real-analytic.
\end{lem}

\begin{proof}
For every $\varphi\in C_c^\infty(U)$, Stieltjes inversion gives
\begin{equation}
 \int_U\varphi(E)\,d n(E)
 =\lim_{\epsilon\downarrow0}\frac{1}{\pi}
  \int_U\varphi(E)\Im m_+(E+i\epsilon)\,d E.
\label{test-inversion}
\end{equation}
On a complex neighborhood of $\supp\varphi$, the function in the integrand
equals $m^{\mathrm{ext}}$ for positive imaginary part.  The convergence to
the real axis is therefore uniform, and \eqref{measure-density} follows.
The restriction of the imaginary part of a holomorphic function to the real
axis is real-analytic.
\end{proof}


\begingroup\small
\begin{thebibliography}{99}

% Bibliographic verification: https://www.jstor.org/stable/24714232 ; erratum: https://www.jstor.org/stable/i24712417
\bibitem{AlbeverioEtAl1984}
S. Albeverio, F. Gesztesy, R. H\o egh-Krohn, and W. Kirsch,
``On point interactions in one dimension,''
\textit{J. Operator Theory} {\bfseries 12}, 101--126 (1984); Erratum,
{\bfseries 13}, 419 (1985).

% Bibliographic verification: https://www.ams.org/publications/authors/books/postpub/chel-350
\bibitem{Albeverio2005}
S. Albeverio, F. Gesztesy, R. H\o egh-Krohn, and H. Holden,
\textit{Solvable Models in Quantum Mechanics}, 2nd ed.
(AMS Chelsea, Providence, 2005).

% Bibliographic verification: https://doi.org/10.1016/S0196-8858(82)80016-X
\bibitem{AlbeverioHKM1982}
S. Albeverio, R. H\o egh-Krohn, W. Kirsch, and F. Martinelli,
``The spectrum of the three-dimensional Kronig-Penney model with random
point defects,''
\textit{Adv. Appl. Math.} {\bfseries 3}, 435--440 (1982).

% Bibliographic verification: https://doi.org/10.1063/1.4769219
\bibitem{DrabkinKirschSchulzBaldes2012}
M. Drabkin, W. Kirsch, and H. Schulz-Baldes,
``Transport in the random Kronig-Penney model,''
\textit{J. Math. Phys.} {\bfseries 53}, 122109 (2012).

% Bibliographic verification: https://doi.org/10.1007/978-3-0348-7306-2_16
\bibitem{KirschNitzschner1990}
W. Kirsch and F. Nitzschner,
``Lifshitz-tails and non-Lifshitz-tails for one-dimensional random point
interactions,''
in \textit{Order, Disorder and Chaos in Quantum Systems},
P. Exner and H. Neidhardt, eds.,
Operator Theory: Advances and Applications 46
(Birkh\"auser, Basel, 1990), pp. 171--178.

% Bibliographic verification: https://doi.org/10.1063/5.0002885
\bibitem{HislopKirschKrishna2020}
P. D. Hislop, W. Kirsch, and M. Krishna,
``Eigenvalue statistics for Schr\"odinger operators with random point
interactions on $\mathbb R^d$, $d=1,2,3$,''
\textit{J. Math. Phys.} {\bfseries 61}, 092103 (2020).

% Bibliographic verification: https://doi.org/10.1007/BF01208718
\bibitem{KirschMartinelli1982}
W. Kirsch and F. Martinelli,
``On the spectrum of Schr\"odinger operators with a random potential,''
\textit{Commun. Math. Phys.} {\bfseries 85}, 329--350 (1982).

% Bibliographic verification: https://doi.org/10.1007/BF01222516
\bibitem{Pastur1980}
L. A. Pastur,
``Spectral properties of disordered systems in the one-body approximation,''
\textit{Commun. Math. Phys.} {\bfseries 75}, 179--196 (1980).

% Bibliographic verification: https://doi.org/10.1070/RM1979v034n02ABEH002908
\bibitem{Shubin1979}
M. A. Shubin,
``The spectral theory and the index of elliptic operators with almost periodic
coefficients,''
\textit{Russ. Math. Surv.} {\bfseries 34}(2), 109--157 (1979).

% Bibliographic verification: https://doi.org/10.1007/BF01205503
\bibitem{CraigSimon1983}
W. Craig and B. Simon,
``Log H\"older continuity of the integrated density of states for stochastic
Jacobi matrices,''
\textit{Commun. Math. Phys.} {\bfseries 90}, 207--218 (1983).

% Bibliographic verification: https://doi.org/10.1007/BF01211591
\bibitem{CampaninoKlein1986}
M. Campanino and A. Klein,
``A supersymmetric transfer matrix and differentiability of the density of
states in the one-dimensional Anderson model,''
\textit{Commun. Math. Phys.} {\bfseries 104}, 227--241 (1986).

% Bibliographic verification: https://doi.org/10.1007/s00023-006-0297-1
\bibitem{BellissardHislop2007}
J. V. Bellissard and P. D. Hislop,
``Smoothness of correlations in the Anderson model at strong disorder,''
\textit{Ann. Henri Poincar\'e} {\bfseries 8}, 1--26 (2007).

% Bibliographic verification: https://doi.org/10.1007/s10955-012-0603-x
\bibitem{KaminagaKrishnaNakamura2012}
M. Kaminaga, M. Krishna, and S. Nakamura,
``A note on the analyticity of density of states,''
\textit{J. Stat. Phys.} {\bfseries 149}, 496--504 (2012).

\end{thebibliography}
\endgroup
\end{document}